%% file: ICEShort15.tex
\documentclass[copyright,creativecommons]{eptcs}
\providecommand{\event}{ICE 2010} 

\providecommand{\firstpage}{1}

\usepackage{breakurl}        
\usepackage{color}
\usepackage{graphicx}
\usepackage{amssymb}
\usepackage{relsize}
\usepackage{xspace}
\usepackage{subfigure}
\usepackage{amssymb,amsbsy,latexsym,amsmath, euscript,amsfonts}
\usepackage{epic,eepic}
\usepackage[all]{xypic}
\usepackage[english]{babel}
\usepackage{epsfig}
\usepackage{url}
\usepackage{dsfont}
\usepackage{vaucanson-g}

 \title{Safer in the Clouds \\(Extended Abstract)
 \thanks{Research supported by the Italian PRIN Project ``SOFT''. }}

\author{Chiara Bodei,
Viet Dung Dinh
and
Gian Luigi Ferrari
 \institute{Dipartimento di Informatica,  Universit\`a di Pisa, Italy}
 \email{\{chiara,dinh,giangi\}@di.unipi.it}
 }
 \def\titlerunning{Safer in the Clouds}
\def\authorrunning{C. Bodei, V.D. Dinh, \& G.L. Ferrari }
\begin{document}
\maketitle

\begin{abstract}
We outline the design of a framework for modelling 
cloud computing systems.The approach is based on a declarative programming model which takes the form of a $\lambda$-calculus enriched with suitable mechanisms to express and enforce application-level security policies governing usages of resources available in the clouds. 
We will focus on the server side of cloud systems, by adopting a {\em pro-active} approach, where explicit security policies regulate server's behaviour. 
\end{abstract}
\input{abstract}

\scriptsize
\bibliographystyle{eptcs}
\bibliography{biblio}

\end{document}

%% file: abstract.tex
\section{Overview}

In the old times, people used to exploit the bakery's oven for their home-made bread. 
Similarly, people utilised the public mill to obtain flour from their wheat.
In both cases, people did not own the physical infrastructure to process their products, neither they invested on it. 
Instead, they rented usage from a third-party provider.
Under this regard, the idea of cloud computing is not completely new, it just updates the above model and adapt it to the Internet world.
Cloud Computing customers do not invest on hardware, software or services, but they just pay providers to use them, either on a utility or a subscription basis.
They rely on clouds for infrastructures, platforms, and software.
Cloud services and the resources they offer over the Internet are therefore 
used on demand and with a certain degree of flexibility.
Thus, Cloud Computing is changing the fundamental way in which computing services are being delivered.

Usually, Cloud services are based on (a farm of) servers, often virtual ones, and are fully managed by their providers.
Consequently, old and new security problems may arise, 
because security is related to trust as more and more customers use Cloud services. 
Since Cloud environment is dynamically changing and shared by a number of users, customers expect their data or resources, when are outsourced, to be protected in Clouds. 
These expectations or requirements are usually captured in Service Level Agreement, and therefore
enforcing SLAs is one of key aspects of Cloud Computing. 
As a result, a number of challenging issues comes into consideration. 
By going deep into it, Cloud services consist of two major parts: business logic and operation logic.
The former models the core functionalities of the services. 
The later formalizes instead the operational tasks, i.e.~the resource behavior of Cloud services. 
Furthermore, from the resource viewpoint, cloud resources come from all levels of the computing stack, 
hence the effective management of the operations acting over dynamically available resources must be handled more carefully, possibly through strict security policies.

Although Cloud Computing has been introduced in order to outsource
IT facilities across the entire stack, integration is one of its important aspects. 
The purpose of integration is to allow customers to create and/or custom applications specifically suited to their unique business processes and policies. 
Hence it gives an ability to modify cloud services to match the needs of a particular goal. 
Again, security issues become more problematic under these circumstances.




In this paper, we advocate the usage of 
a declarative model based on functions with side effect to abstractly represent services acting over resources and service assemblies.
Our main contribution is the formal semantics of cloud servers as functions, expressed in a suitable dialect of the $\lambda$-calculus
and the modelling of clients interactions is via asynchronous invocation.
Next, we informally present the main features of our approach.

\noindent
{\bf Services as functions with side effects} 
We adopt the idea of \emph{Software as a service}: each service exposes over the network certain functional behaviour and it is invoked via request/response communication protocols (e.g.~SOAP). 
Services are pluggable entities and composite services are obtained by combining existing elementary or complex services.
Following \cite{BartolettiDFZ08,Bartoletti09Toplas} cloud services are viewed as functions. However, a service is not a \emph{pure} function as its execution yields a side effect, thus reflecting changes of the service state. 
For instance, let us consider a simple storage service, offered by a cloud, that gets a string query from the
user and accordingly queries the database. 
The following functional interface can be used to describe the above service.

\begin{center}
\texttt{Table fun Q(Query q): Effect e}
\end{center}
\noindent
The invocation of the service \texttt{Q} with a query value will yield a table value as result. 
The side effect \texttt{e} provides the abstract representation of the modifications to the database such as updates of tables. 
Here, the side effects represent the action of accessing service resources.

The main benefit of the service-as-function metaphor, with respect to alternative approaches based on process calculi 
(see~\cite{Bruni09} for a survey), 
is that it provides a \emph{high-level} notion 
to model services, their composition and interactions, by abstracting from low level details.
%

\noindent
{\bf Security Policies}
In spite of its undeniable advantages and cost-savings, 
cloud computing 
makes data processing inherently risky, as
data reside not under the user's control.
Consequently, it is crucial that safety is addressed when designing a cloud.

Our programming model focuses on application-based security by considering \emph{security policies} as first class programming constructs. 
We provide explicit constructs to declare and enforce the security policies governing the behaviour of applications, 
in the style of \cite{BartolettDF05,BDFZ07}. A security policy regulates
how resources are granted to and used by services.
For instance, let us consider the database service example above. 
The \texttt{Q} service may be unsafe although the code normally runs in most of the cases. 
An attacker can indeed taint the query string by injecting a command in front ({\em SQL injection bug});
consequently the service would issue dangerous commands such as deleting a file before
executing the safe query. 


From a methodological perspective, being security-oriented from the very beginning of the development process facilitates the design of secure services: security is faced in advance, without sweeping it under the carpet (read it as security patches added later).
The database service example above can be moved into a more secure land, by wrapping it inside a suitable security policy 
$\varphi_{DB}$. For instance, the policy can impose that no update operations on the database (i.e.~system commands) can be issued during service executions, i.e.~the only operations allowed are those in which the database content can only be read.
Adding the specified policy to the query interface results in:
%
\begin{center}
\texttt{Table fun Q(Query q): Effect e ensuring} $\varphi_{DB}$
\end{center}
\noindent
meaning that each step of service execution must obey the security policy $\varphi_{DB}$. 
Operationally, the run-time structures will enforce the security policy $\varphi_{DB}$
by monitoring service execution and by catching the occurrences of bad actions, i.e.~the actions that violate the policies.  
Actually, the run-time enforcement mechanism depends on
a suitable abstraction of the execution of all the pieces of code 
(possibly partially) executed so far. 
This implies that the mechanism enforcing the security policies can make decisions, 
based on all previous changes of shared resources affected by different user requests.
This approach, known under the name of \emph{history-based security}, has been receiving major attention, at both levels of
foundations~\cite{Banerjee04history,Fong04shallow,Skalka04history}
and of language design/implementation~\cite{Abadi03history,Edjlali99history}. 

\noindent
{\bf Cloud server}
Abstractly a cloud server can take the form of a pool of services and computational resources running over a variety of virtual machines. 
More precisely, a cloud server is a triple 
$(\eta \rhd e_1 \parallel e_2 \dots \parallel e_k \lhd \sigma)$. The pair $(\eta, \sigma)$ is the global cloud state. 
The $\eta$-component, called \emph{history}, stores the abstract representation of the cloud behaviour. More precisely, this component exposes the evolution of the cloud, by detailing the occurrences of the concrete bindings among resources and services. 
The $\sigma$-component, called \emph{service store}, provides a virtual representation of the service metadata supporting facilities to store and retrieve services. 
Here, a service store $\sigma$ is  a function mapping each service name $p_i$ with the metadata and scripts used to load the virtual machine and the resources required to run the service. Intuitively, the service name plays the role of of the service functional interface and our service store is a sort of service directory.
Our semantic framework handles service metadata as \textit{persistent} entities, i.e.~they are not consumed by an invocation, but they remain in the service store.
Finally, the last component 
$e_1 \parallel e_2 \dots \parallel e_k$ represents the expression that describes the set of services running in parallel. 
Services are functions similar to expressions in the
$\lambda$-calculus with a side effect acting on the global cloud state. Services are equipped with suitable primitives for accessing resources, for declaring and enforcing security policies, and for installing services and managing their invocation. For simplicity, we assume that resources are objects that are already available in the cloud environment (i.e.~resources cannot be dynamically created).

We can specify the database service discussed above as follows:\\
\[\begin{array}{l}
E_{form} = \lambda{}q.\ E_{proc}\ q 
\ \  \mbox{ where } \ \
E_{proc} = \lambda{}y.\ open(db); (query\ db\ y); close(db),
\end{array}\]
where $E_{form} $ is the cloud service acting as logical front-end with the actual database. 
The service gets a query string $\langle q \rangle$ from a user, then feeds it to $ E_{proc} $.  
In turn, the function $ E_{proc} $ takes the query $ y $ as a parameter, connects to the database $ db $, makes a query to $ db $, by exploiting an auxiliary function $ query $ with the database and the query string as parameters, then closes the database connection. 
We abstract from the details of the code of  the $ query $ function here, we just assume that it may perform some internal activities and then issues the database command $ dbcmd $ and returns a value $ v $. 

Assume now that the configuration of the cloud server takes the form
$\eta \rhd \mbox{link} \: E_{form} \parallel E \lhd \sigma$. Here, the cloud server has activated the book-keeping service $\mbox{link } E_{form}$ that in parallel with the execution of the services $E$ performs the dynamic publication of the database service $E_{form}$. 
The constructor $\mbox{link} \: E$ is a novel construct of the calculus introduced to describe the dynamic publication of a service whose code is $ E $. Operationally, the execution of the link-construct has the side effect of installing in the service store the service $E$ including all the required metadata.

In our running example, this leads to the following evolution of the cloud server:
\[(\eta \rhd \mbox{link} \: E_{form} \parallel E \lhd \sigma)
\xrightarrow{link}
(\eta \cdot \mbox{link} \rhd E \lhd \sigma[Q \rightarrow E_{form}])\]
\noindent
The target configuration indicates that a new service, called $Q$, is now available in the cloud. Furthermore, the history $\eta$ logs the fact that a dynamic publication has been performed. 

Client interactions are modelled \emph{explicitly}: they occur via asynchronous communication in the style of the asynchronous $\pi$-calculus. Intuitively, the invocation of the service $ \pi $ available in the service storage is only observed by the asynchronous occurrence of the activation of the code associated with $ \pi $. For instance, in our running example, we have:
\[(\eta \cdot \mbox{link} \rhd E \lhd \sigma[Q \rightarrow E_{form}])
\xrightarrow{invoke_{Q}(q)}
(\eta \cdot \mbox{link} \cdot \mbox{invoke}_Q(q) \rhd   E_{form} q \parallel E \lhd \sigma[Q \rightarrow E_{form}]).\]
\noindent
This transition, that uses the new construct $\mbox{invoke}_Q (q)$,
indicates that an instance of the service $Q$ has been activated on the actual parameter $q$. This  instance of the service runs in parallel with the other active services. Moreover, the service $Q$ is still available in the cloud (i.e.~services are permanent objects) and the history $\eta$ logs the fact that an activation of the service has been performed. 

Our treatment of service invocation has the consequent benefit to
manage a variety of clients, by abstracting from the specific interaction protocols
established with the cloud.

Back to our running example, the database service runs as follows, where $ \eta' $ stands for $ \eta \cdot \mbox{link} \cdot \mbox{invoke}_Q(q) $:\\
\[\begin{array}{l}
\ \ \ \ \ \ \ 
( \eta' \rhd  E_{form} q \parallel E \lhd \sigma[Q \rightarrow E_{form}]) 
\\
\xrightarrow{\tau}\ 
(\eta', \rhd E_{proc}\ q \parallel E 
\lhd \sigma[Q \rightarrow E_{form}])
\\
\xrightarrow{\tau}\ 
(\eta', \rhd open(db);((query\ db\ q); close(db)) \parallel E 
\lhd \sigma[Q \rightarrow E_{form}])
\\
\xrightarrow{open(db)}\ 
(\eta' \cdot open(db)  \rhd ((query\ db\ q); close(db)) \parallel E 
\lhd \sigma[Q \rightarrow E_{form}])
\\
\xrightarrow{\tau}^*\xrightarrow{dbcmd}\xrightarrow{\tau}^*\ 
(\eta' \cdot open(db) \cdot dbcmd \rhd close(db) \parallel E \lhd
\sigma[Q \rightarrow E_{form}])
\\
\xrightarrow{\tau}^*\xrightarrow{close(db)}\ 
(\eta' \cdot open(db) \cdot dbcmd \cdot close(db) \rhd  E \lhd 
\sigma[Q \rightarrow E_{form}])
\end{array}\]

As previously discussed, this behaviour is unsafe because it may contain a SQL injection bug: an attacker can try to inject a command in front of query string, e.g.~$syscmd;q'$, and therefore can execute any dangerous command such as deleting a file as as illustrated by the following
trace
:
\[\begin{array}{l}
\ \ \ \ \ \ \ 
\vdots
\\
\xrightarrow{\tau}\ 
(\eta', \rhd open(db);((query\ db\ (syscmd;q')); close(db)) \parallel E 
\lhd \sigma[Q \rightarrow E_{form}])
\\
\xrightarrow{open(db)}\ 
(\eta' \cdot open(db)  \rhd ((query\ db\ (syscmd;q')); close(db)) \parallel E 
\lhd \sigma[Q \rightarrow E_{form}])
\\
\xrightarrow{syscmd}\ 
(\eta' \cdot open(db) \cdot syscmd \rhd ((query\ db\ q'); close(db)) \parallel E 
\lhd \sigma[Q \rightarrow E_{form}])
\\
\xrightarrow{\tau}^*\xrightarrow{dbcmd}\xrightarrow{\tau}^*\ 
(\eta' \cdot open(db) \cdot syscmd \cdot dbcmd \rhd close(db) \parallel E \lhd
\sigma[Q \rightarrow E_{form}])
\\
\xrightarrow{\tau}^*\xrightarrow{close(db)}\ 
(\eta' \cdot open(db) \cdot syscmd \cdot dbcmd \cdot close(db) \rhd  E \lhd 
\sigma[Q \rightarrow E_{form}])
\end{array}\]

To prevent this, we instrument the code $E_{form} $ by framing it with a security policy $\varphi_{DB} $, which does not allow execution of any \textit{system command}. 

\begin{figure}
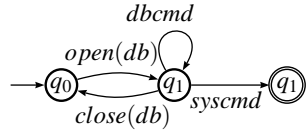

\centering
\SmallPicture\VCDraw{%
\begin{VCPicture}{(0,0)(8,4)}
\State[q_0]{(1,2)}{A} \State[q_1]{(4,2)}{B} \FinalState[q_1]{(7,2)}{C}
\Initial{A}
\ArcL{A}{B}{}  \LabelL{open(db)}
\ArcL{B}{A}{}  \LabelL{close(db)}
\LoopN{B}{} \LabelL{dbcmd}
\EdgeL{B}{C}{}  \LabelR{syscmd}
\end{VCPicture}
}
\caption{Usage automaton of the service \texttt{Q}}. \label{fig:sqlservice}
\end{figure}

Following the approach of~\cite{BartolettDF05,BDFZ07}, a security policy is a set of traces that describe the sequences of events satisfying that policy. 
Security policies are specified via \emph{usage automata} \cite{Bartoletti09Wits} a suitable extension of Finite State Automata (FSA),
whose final states denote policy violations. The usage automation of $ \varphi_{DB} $ is depicted in Fig.\ref{fig:sqlservice} and the following traces show 
how to prevent execution of any system command:
\[\begin{array}{l}
\ \ \ \ \ \ \ 
\vdots
\\
\xrightarrow{open(db)}\ 
(\eta' \cdot open(db)  \rhd \varphi_{DB}[((query\ db\ (syscmd;q')); close(db))] \parallel E 
\lhd \sigma[Q \rightarrow E_{form}])
\\
\not\xrightarrow{syscmd}\ 
\\
\end{array}\]

%
%
%
%
%
%
%
%

In the full paper \cite{TR}, we introduced a notion of abstract semantics for cloud server by extending the notion of \emph{applicative bisimulation}~\cite{Abramsky93}. 
Intuitively, the idea behind applicative bisimulation is that in order to reason on the equivalence of two functions, we need to know whether their behaviors are the same with all possible closed values. As a consequence, applicative bisimulation relies on the output generated by functions, hence it does not capture the events issued by our functions The management of events is crucial in our framework.  Service behaviour is indeed \emph{history-dependent}, i.e.~an expression may be executed differently when plugged within different cloud states. 
For instance, let us consider the cloud servers:
$ (\eta \rhd \beta;\alpha;\varphi[\gamma] \lhd \sigma) $ and $ (\eta \rhd \alpha;\beta;\varphi[\gamma] \lhd \sigma) $, where $ \eta $ contains neither $ \alpha $ nor $ \beta $, and the policy $\varphi$ states that the sequence $ \alpha;\beta $ is \textit{not} allowed. After two transitions, the first configuration can make a transition that issues $ \gamma $, while the second cannot.
We obtain a particular bisimulation (called {\em cloud bisimulation}) that is proved to be also a congruence and that allows us to prove that
policy framings do not add new behaviors to the wrapped programs, i.e.~behaviours of the wrapped programs are always simulated by their origins.  
Intuitively, behaviour that violates the policy is prevented to occur. 
In the running example, it appears that $ \varphi_{DB}[E_{form}] \lesssim E_{form} $.

In summary, technical results give rise to a semantics-based methodology of safety refinement process in cycle of software development.
Even though our present definition of bisimilarity would require an infinite number of checks, we are confident to circumvent
this problem, by resorting to symbolic techniques, like the one defined in the field of software model checking.